\documentclass[journal]{IEEEtran}
\usepackage{algorithm}
\usepackage{algpseudocode}
\usepackage{amsfonts}
\usepackage{bm}
\usepackage{amsmath}
\usepackage{cite}
\usepackage{float}
\usepackage{amssymb}
\usepackage[timestamp,first]{draftcopy}
\usepackage{enumerate}
\usepackage{color,xcolor}
\usepackage{colortbl}
\usepackage{graphicx}
\usepackage{multirow,tabularx}
\usepackage{subfigure}

\makeatletter

\newcommand{\Rmnum}[1]{\expandafter\@slowromancap\romannumeral #1@}
\makeatother
\usepackage{bm}

\ifCLASSINFOpdf

\else

\fi

\begin{document}

\title{Message Passing in C-RAN: Joint User Activity and Signal Detection}

\author{\IEEEauthorblockN{Yuhao Chi\IEEEauthorrefmark{1}, Lei Liu\IEEEauthorrefmark{2}\IEEEauthorrefmark{3}, Guanghui Song\IEEEauthorrefmark{4}, Chau Yuen\IEEEauthorrefmark{2}, Yong Liang Guan\IEEEauthorrefmark{6}, and Ying Li\IEEEauthorrefmark{1}}\\
\IEEEauthorrefmark{1}State Key Lab of ISN, Xidian University, China,
\IEEEauthorrefmark{2}Singapore University of Technology and Design, Singapore,
\IEEEauthorrefmark{3}City University of Hong Kong, China,
\IEEEauthorrefmark{4}Doshisha University, Kyoto, Japan,\\
\IEEEauthorrefmark{6}Nanyang Technological University, Singapore
\thanks{This work was supported in part by the National Natural Science Foundation of China under Grants 61671345, in part by the Singapore A*STAR SERC Project under Grant 142 02 00043, in part by the Japan Society for the Promotion of Science through the Grant-in-Aid for Scientific Research (C) under Grant 16K06373, and in part by the Ministry of Education, Culture, Sports, Science and Technology through the Strategic Research Foundation at Private Universities (2014-2018) under Grant S1411030. The first author was also supported by the China Scholarship Council under Grant 201606960042.
}
}

\maketitle

\begin{abstract}
In cloud radio access network (C-RAN), remote radio heads (RRHs) and users are uniformly distributed in a large area such that the channel matrix can be
considered as sparse. Based on this phenomenon, RRHs only need to detect the relatively strong signals from nearby users and ignore the weak signals from far users, which is helpful to develop low-complexity detection algorithms without causing much performance loss. However, before detection, RRHs require to obtain the real-time user activity information
by the dynamic grant procedure, which causes the enormous latency. To address this issue, in this paper, we consider a grant-free C-RAN system and propose a low-complexity Bernoulli-Gaussian message passing (BGMP) algorithm based on the sparsified channel, which jointly detects the user activity and signal. Since
active users are assumed to transmit Gaussian signals at any time, the user activity can be regarded as a Bernoulli variable and the signals from all users obey a Bernoulli-Gaussian distribution. In the BGMP, the detection functions for signals are designed with respect to the Bernoulli-Gaussian variable.
Numerical results demonstrate the robustness and effectivity of the BGMP. That is,
for different sparsified channels, the BGMP can approach the mean-square error (MSE) of the genie-aided sparse minimum mean-square error (GA-SMMSE) which
exactly knows the user activity information. Meanwhile, the fast convergence and strong recovery capability for user activity of the BGMP are also verified.
\end{abstract}

\IEEEpeerreviewmaketitle
\section{Introduction}
To support massive data demands in wireless communications, cloud radio access network (C-RAN) emerges as a candidate for the next generation network architecture, which can significantly improve spectral efficiency and energy efficiency~\cite{C-ran,FanMaga,Zuo}. Unlike traditional multiuser multiple-input multiple-output (MU-MIMO) systems, C-RAN consists of hundreds of remote radio heads (RRHs) deployed in a large area and a pool of baseband units (BBUs)
centralized in a data cloud center. All RRHs collect signals from users and merge all signals to BBUs for signal recovery.

In order to reliably recover signals with low complexity, a promising detection method is the message passing algorithm based on factor graph~\cite{Loeliger2006,Donoho2009,GAMP,Chongwen}, which transforms the optimal cost function for signal recovery into the iterative calculations among nodes in the factor graph. For different networks, the message passing algorithm need to be specially designed. In C-RAN, affected by the path loss, the signals from far users are very weak when arrive at RRHs, which results in a nearly sparse channel. Authors in~\cite{Fan_Dynamic2016} proved that the C-RAN channel could be sparsified without causing much performance loss, where RRHs only needed to detect the relatively strong signals from nearby users and ignored the weak signals from far users. The channel sparsification is helpful to develop low-complexity message passing algorithms. Nevertheless, due to different statistic distributions of channels, the Gaussian message passing (GMP) algorithms proposed for the MU-MIMO~\cite{Lei2015,Lei2016,Lei20162} cannot be directly extended to the C-RAN. As a result, authors in \cite{Fan2015,Fan2016} proposed a
sparse message passing algorithm for the C-RAN with the sparsified channel~\cite{Fan_Dynamic2016}.

However, in the above works~\cite{Lei2015,Lei2016,Lei20162,Fan_Dynamic2016,Fan2015,Fan2016}, receivers are assumed to exactly know the real-time user activity information and then detect the signals for active users. In practice, the user activity information is obtained by the complicated grant procedure. When the number of users is large and the activity of each user changes at any time, the dynamic grant procedure causes the enormous latency. To address this issue, authors in~\cite{XXu2015,Utkovski2017} considered a grant-free C-RAN system and proposed a modified Bayesian compressive sensing and a hybrid generalized approximate message passing (GAMP) respectively, which were to estimate the channel state information and user activity. However, in~\cite{XXu2015,Utkovski2017}, the channel models do not take into account the geographical distributions of RRHs and users and these algorithms do not consider the signal recovery for active users.

In this paper, we consider joint user activity and signal detection over the grant-free C-RAN. Since the activity of each user changes at any time,
the user activity can be regarded as a Bernoulli variable at RRHs. Moreover, we assume that active users transmit Gaussian signals and
the transmissions of inactive users can be treated as zeros for RRHs. Statistically, the signals from all users obey a Bernoulli-Gaussian distribution.
Therefore, based on the sparsified channel and corresponding factor graph, we propose a low-complexity Bernoulli-Gaussian message passing (BGMP) algorithm to
 jointly detect the user activity and signal. In the BGMP, messages passing among nodes and relevant update functions at nodes are associated with the Bernoulli-Gaussian variable. Numerical results demonstrate the robustness and effectivity of the BGMP. That is, for different sparsified channels, the BGMP can approach the mean-square error (MSE) of the genie-aided sparse minimum mean-square error (GA-SMMSE) which exactly knows the user activity information. Moreover, the fast convergence and strong recovery capability for user activity of the BGMP are also verified.
\section{System Model}
\begin{figure}[h!] \vspace{-0.3cm}
\centering
\includegraphics[width=0.8\columnwidth]{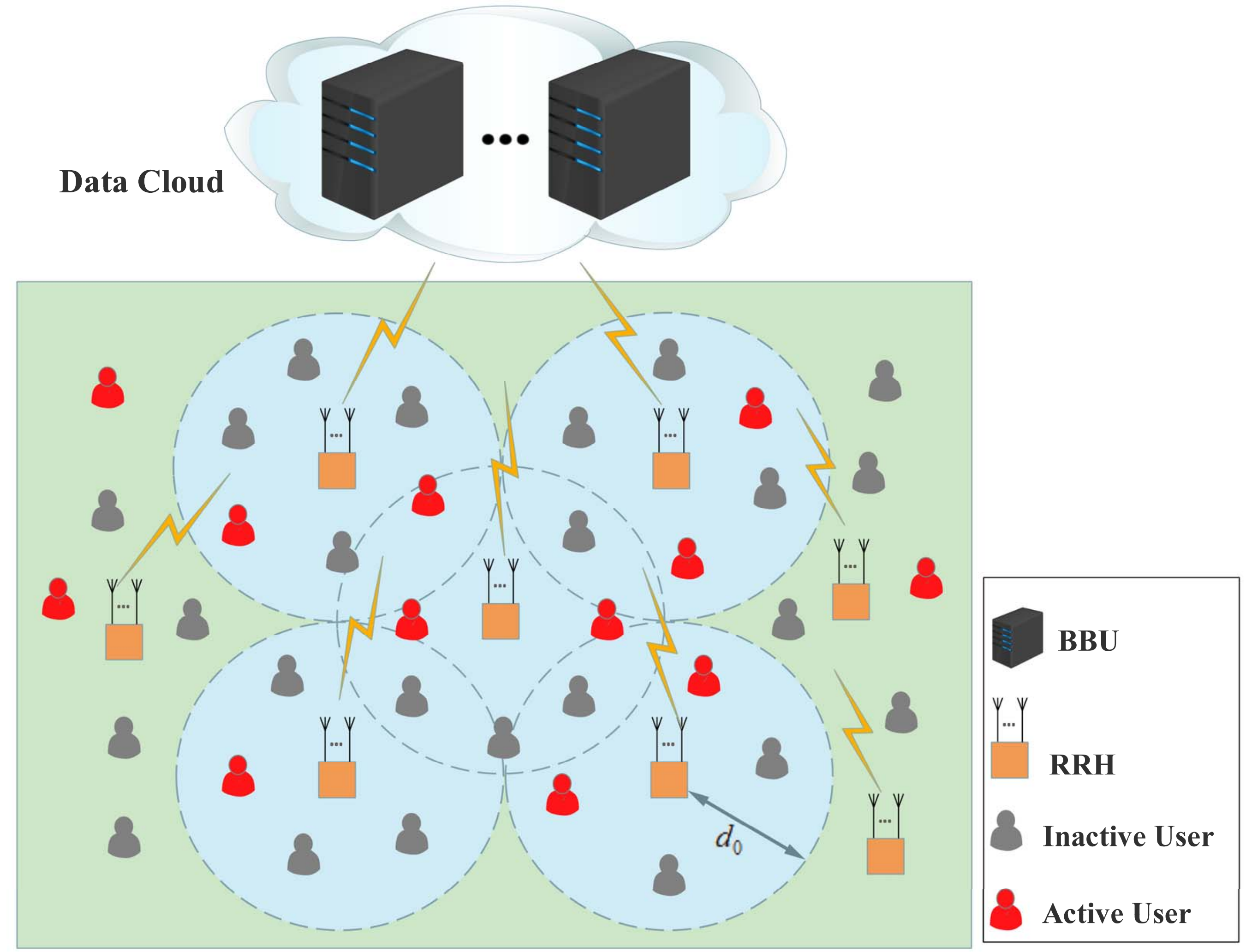}\\
\caption{Illustration of an uplink grant-free C-RAN system, where RRHs and users are uniformly located over a large coverage area.}\label{Model}
\end{figure}\vspace{-0.1cm}
Figure \ref{Model} shows an uplink grant-free C-RAN system with $M$ RRHs and $K$ users uniformly located over a large coverage area, where active users transmit signals at any time without the complicated grant procedure. Each RRH has $N$ antennas and each user has one antenna. Signal ${{\bm{y}}^m} \in {\mathcal{R}}^{{N} \times {\rm{1}}}$ arrived at the $m$-th RRH is 
\begin{equation}\label{rev}
{{\bm{y}}^m}=P^{\frac{1}{2}}{\bm{H}}^m {\bm{x}}+{\bm{z}}^m,\quad m=1, ..., M 
\end{equation}
where ${\bm{H}}^m$ $\in$ ${\mathcal{R}}^{N \times K}$ denotes the channel matrix from $K$ users to the $m$-th RRH, $P$ is the transmit power allocated to each user, ${\bm{x}}$ $\in$ ${\mathcal{R}}^{K \times {\rm{1}}}$ is the transmitted signal from $K$ users, and ${\bm{z}}^m$$\in$ ${\mathcal{R}}^{N \times {\rm{1}}}$ is a Gaussian noise vector obeying $\mathcal{N}(0, \sigma_{n}^{2}\bm{I}_{N})$ with an $N\times N$ identity matrix $\bm{I}_{N}$. The~$(n, k)$-th entry $h^m_{n,k}$ of ${\bm{H}}^m$ is assumed as $\gamma^m_{n,k}d_{m,k}^{-{\alpha}}$, where $\gamma^m_{n,k}$ is an independent and identically
distributed (i.i.d.) fading coefficient obeying ${\mathcal{N}}(0,1/K)$, $d_{{m,k}}$ is the geographic distance between the $k$-th user and the~$m$-th RRH, and $\alpha$ is a path loss exponent. Note that $d_{m,k}^{-{\alpha}}$ denotes the path loss from the $k$-th user to the $m$-th RRH. Here, we assume that the $m$-th RRH perfectly knows channel state information ${\bm{H}}^m$.

Due to the effect of path loss, the received signals from far users are drastically degraded such that RRHs can ignore the detections for far users without
causing much performance loss~\cite{Fan_Dynamic2016}. The channel sparsification can provide a sparse factor graph to develop low-complexity message passing
algorithms~\cite{Fan2015,Fan2016}. Therefore, as~\cite{Fan2015,Fan2016}, we set a distance threshold $d_0$ to sparsify the channel in~Fig.~\ref{Model}.
Specifically, the $(n, k)$-th entry ${\hat{h}}^{m}_{n,k}$ of sparsified channel
matrix ${\bm{\hat{H}}}^m$ is
\begin{equation}\nonumber
{\hat{h}}^{m}_{n,k} = \left\{
{\begin{array}{*{20}c}
\!\!\!{h}^m_{n,k},\quad \;{d_{m,k} < d_0}, \\
{0,\quad \quad\rm{otherwise}.}
\end{array}} \right.
\end{equation}
Then, Eq.~(\ref{rev}) is rewritten as
\begin{align} \nonumber
{\bm{y}}^m&=P^{\frac{1}{2}}{\bm{\hat{H}}}^m{\bm{x}}+P^{\frac{1}{2}}{\bm{\tilde{H}}}^m{\bm{x}}+{\bm{z}}^m \\ \label{newRev}
&=P^{\frac{1}{2}}{\bm{\hat{H}}}^m{\bm{x}}+\bm{\eta}^m\rm{,}
\end{align}
where ${\bm{\tilde{H}}}^m$$=$${\bm{H}}^m$$-{\bm{\hat{H}}}^m$ and $\bm{\eta}^m$$=$$P^{\frac{1}{2}}$${\bm{\tilde{H}}}^m$${\bm{x}}$$+{\bm{z}}^m$ is an interference
vector of length $N$. The variance of the $n$-th entry $\eta^m_n$ of $\bm{\eta}^m$ is ${\sigma}^2_{mn}=PE[\sum_{k}|\tilde{h}^m_{n,k}x_k|^2]+\sigma_{n}^{2}$,
where $\tilde{h}^m_{n,k}$ is the ($n$, $k$)-th entry of ${\bm{\tilde{H}}}^m$, $x_k$ is the $k$-th entry of $\bm{x}$, $k=1, ..., K$ and $n=1, ..., N$.

Note that in the grant-free C-RAN, RRHs cannot obtain the user activity information in advance. Thus, the user activity can be regarded as a
Bernoulli variable at the RRHs. Moreover, we assume that active users transmit Gaussian signals. The transmissions of inactive users can be treated as
zeros for RRHs. Statistically,
entries of ${\bm{x}}$ obey a Bernoulli-Gaussian distribution. That is,
\begin{equation}\nonumber
x_k = \left\{
{\begin{array}{*{20}c}
{0,\quad \quad \quad\quad\;{\rm{with~probability}}\; 1-\rho, } \\
{\!\!\!\!\!\!\!\!\!\!\!{\mathcal{N}}(0,\rho^{-1} ),\;\;\;{\rm{with~probability }}\;\rho },
\end{array}} \right.
\end{equation}
where $0 < \rho < 1$ is the probability of user activity and the power of $x_k$ is normalized to $1$. Our goal is to develop a low-complexity message
passing algorithm based on the sparsified channel, which jointly detects the user activity and signal.
\section{Bernoulli-Gaussian Message Passing}
\begin{figure}[!h]\vspace{-0.2cm}
\centering
\includegraphics[width=1\columnwidth]{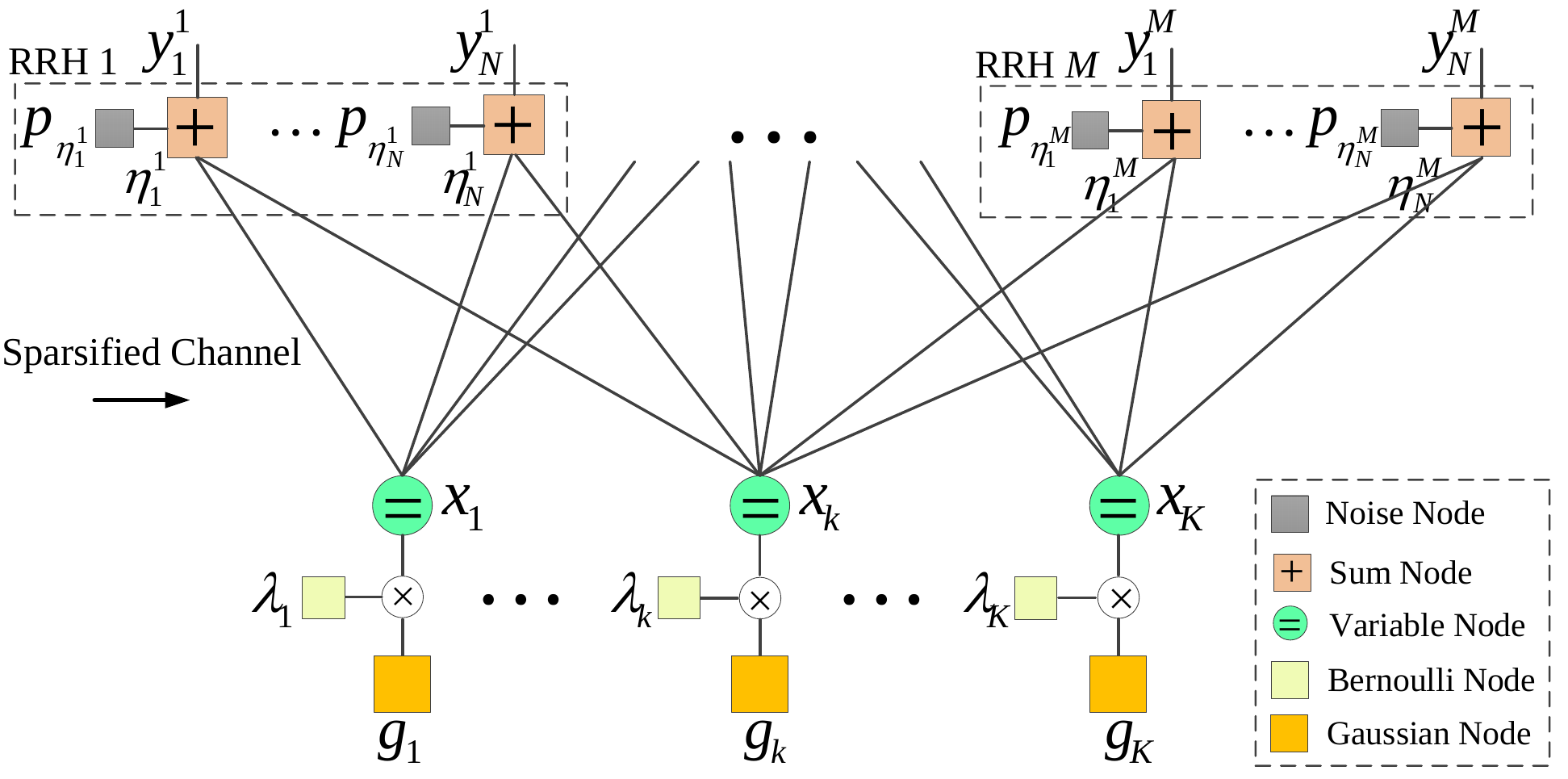}\\ \vspace{-0.2cm}
\caption{Factor graph of the grant-free C-RAN with the sparsified channel. There are $M$ multiple-antenna RRHs in which each RRH has $N$ antennas denoted as sum nodes, and $K$ single-antenna users denoted as variable nodes. Bernoulli vector ${\bm{\lambda}}=[{\lambda_1, ..., \lambda_K}]^T$ and Gaussian signal vector ${\bm{g}}=[g_1, ..., g_K]^T$ are denoted as Bernoulli and Gaussian nodes respectively.}\label{FG} 
\end{figure}
\vspace{-0.1cm}
\begin{figure*}[!ht]
\centering
\includegraphics[width=1.3\columnwidth]{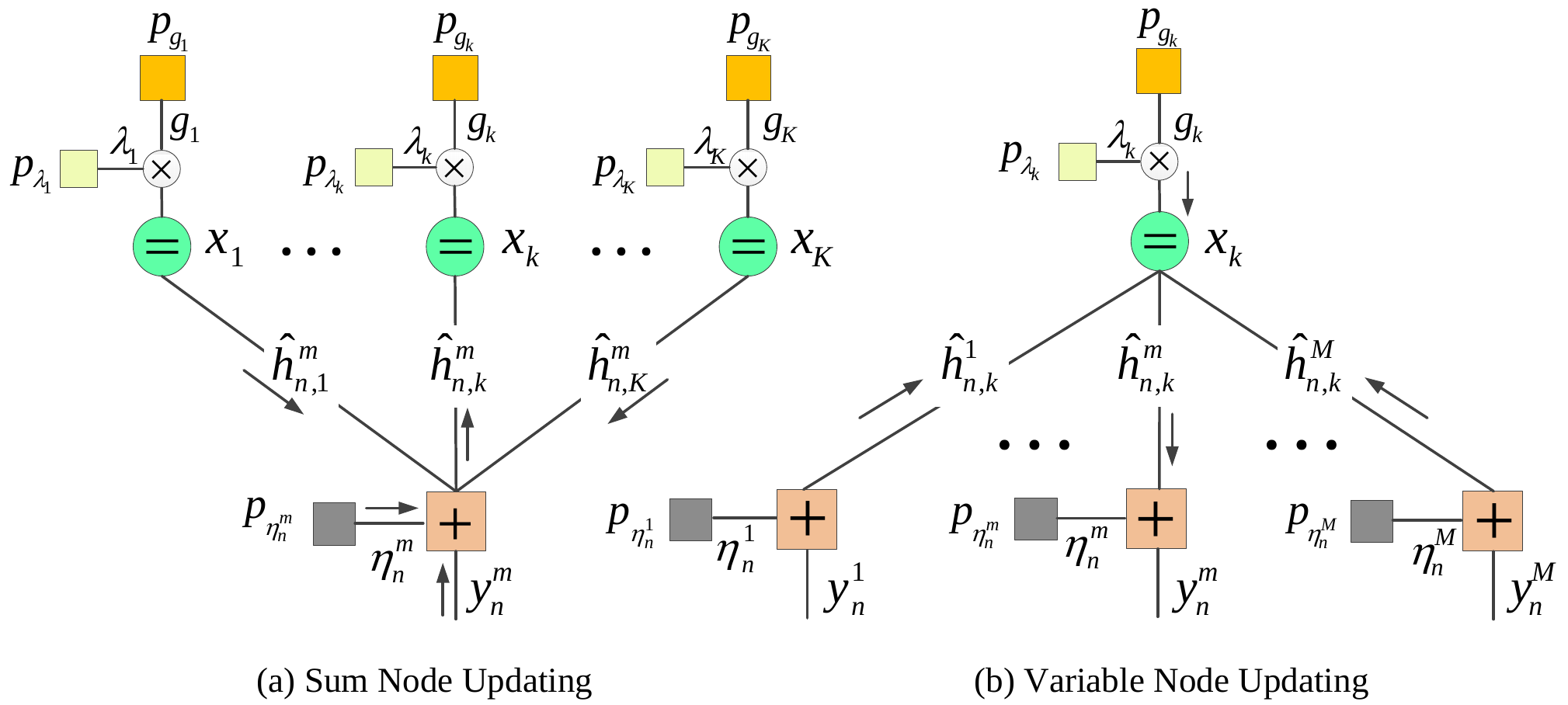}\\ \vspace{-0.2cm}
\caption{Bernoulli-Gaussian message updates between the $n$-th sum node of the $m$-th RRH and the $k$-th variable node including Bernoulli node $\lambda_k$ and Gaussian node $g_k$, $m=1, ..., M$, $n=1, ..., N$, and $k=1, ..., K$. Messages along edges consist of non-zero probability of $\lambda_k$ and mean and variance of~$g_k$.}\label{MessPass} \vspace{-0.4cm}
\end{figure*}
To identify active users and recover their signals, we propose a Bernoulli-Gaussian message passing (BGMP) algorithm for the C-RAN with the sparsified channel.
To simplify the analysis, Bernoulli-Gaussian vector $\bm{x}$ is transformed into the componentwise product of a Bernoulli vector ${\bm{\lambda}}$ obeying i.i.d. $\mathcal{B}(1,\rho\bm{I}_K)$ and a Gaussian vector ${\bm{g}}$ obeying i.i.d.~$N(0, \rho^{-1}\bm{I}_K)$, i.e.,\vspace{-1mm}
\[
{\bm{x}}={\bm{\lambda}} \circ \bm{g},
\vspace{-1mm}
\]
where ${\bm{\lambda}}$ and ${\bm{g}}$ are independent of each other and $\circ$ refers to the element-wise multiplication. Thus, the recovery for $\bm{x}$ is transformed  into the joint recovery for $\bm{\lambda}$ and $\bm{g}$. Fig.~\ref{FG} shows the factor graph of the C-RAN, where antennas of all RRHs, users, $\bm{\lambda}$, and $\bm{g}$ are denoted as sum, variable, Bernoulli, and Gaussian nodes respectively.

As the signal detection in conventional message passing algorithms, such as GMP algorithm~\cite{Lei2015} and belief propagation (BP) decoding of LDPC code~\cite{LDPC}, in the proposed BGMP algorithm, we decompose the global calculation based on the full channel matrix into many local calculations at nodes in the factor graph. This is efficient to reduce the computation complexity. Note that the messages in GMP or BP decoding relate to Gaussian or discrete signals. Different from GMP and BP decoding, the messages in the BGMP are associated with both Bernoulli and Gaussian signals. Fig.~\ref{MessPass} illustrates the update processes of messages between sum nodes and variable nodes in the BGMP algorithm. To be specific, we present message updates at sum and variable nodes as follows.
\subsection{Bernoulli-Gaussian Message Update at Sum Node}
For simplicity, we consider the detection for user~$k$ at the $m$-th RRH, where user $k$ is nearby the $m$-th RRH, i.e., ${\hat{h}}^{m}_{n, k}=1, n=1, ..., N$. Fig.~\ref{MessPass}(a) shows the update process for messages passing from the~$n$-th sum node of~the $m$-th RRH to the~$k$-th variable node. Then, we rewrite Eq.~(\ref{newRev}) as
\begin{align}\nonumber
y^m_n &=\hat{h}^m_{n,k} \lambda_k g_k+ \sum_{i \in \mathcal{K}\setminus k} \hat{h}^m_{n,i} \lambda_i g_i + \eta^m_n \\ \nonumber
&=\hat{h}^m_{n,k} \lambda_k g_k+ {\hat{\eta}}^m_{nk},
\end{align}
where ${\hat{\eta}}^m_{nk}=\sum_{i \in \mathcal{K}\setminus k}\hat{h}^m_{n,i} \lambda_i g_i + \eta^m_n$, $\mathcal{K}=\{1, ..., K\}$,~and~$i \in \mathcal{K} \setminus k$ denotes that $i \in \mathcal{K}$ and $i \neq k$. Due to independent transmissions of all users, based on the central limit theorem, ${\hat{\eta}}^m_{nk}$ can be regarded as a Gaussian variable with mean $e^m_{nk}$ and variance $v^m_{nk}$. At the $t$-th iteration,
\begin{eqnarray}\label{ETAM}
\lefteqn{\!\!\!\!\!\!\!\!\!\!\!\!\!\!\!\!\!\!\!\!\!\!\!\!\!\!\!\!\!\!\!\!\!\!\!\!\!\!\!\!\!\!\!\!\!\!\!\!\!\!\!\!\!\!\!\!\!\!\!\!\!\!\!\!\!\!\!\!\!\!
e^m_{nk}(t)=E[{\hat{\eta}}^m_{nk}(t)]=\sum_{i \in \mathcal{K}\setminus k}\hat{h}^m_{n,i}p^m_{i\rightarrow n}(t) e^m_{i\rightarrow n}(t),}\\
\label{ETAV}
\lefteqn{\!\!\!\!\!\!\!\!\!\!\!\!\!\!\!\!\!\!\!\!\!\!\!\!\!\!\!\!\!\!\!\!\!\!\!\!\!\!\!\!\!\!\!\!\!\!\!\!\!\!\!\!\!\!\!\!\!\!\!\!\!\!\!\!\!\!\!\!\!\!
v^m_{nk}(t)=Var[{\hat{\eta}}^m_{nk}(t)]}
\end{eqnarray}
\begin{equation}\nonumber
\resizebox{1\hsize}{!}{$=\sum_{i \in \mathcal{K}\setminus k}(\hat{h}^m_{n,i})^2 p^m_{i\rightarrow n}(t)\big(v^m_{i\rightarrow n}(t)+
(1-p^m_{i\rightarrow n}(t))e^m_{i\rightarrow n}(t)^2\big) + {\sigma}^2_{mn},$}
\end{equation}
where $E[a]$ and $Var[a]$ denote the expectation and variance of variable $a$, $e^m_{i\rightarrow n}(t)$ and $v^m_{i\rightarrow n}(t)$ are the mean and variance of $g_i$, and $p^m_{i\rightarrow n}(t)$ is the non-zero probability of $\lambda_i$. These input messages associated with $g_i$ and $\lambda_i$ are from the $i$-th variable node. Based on these priori inputs, the $n$-th sum node of the
$m$-th RRH outputs  mean $e^m_{n\rightarrow k}(t)$ and variance $v^m_{n\rightarrow k}(t)$ for $g_k$, and non-zero probability $p^m_{n\rightarrow k}(t)$ for
$\lambda_k$, which are sent to the $k$-th variable node.

\emph{1) Gaussian message update for $g_k$$\sim$${\mathcal{N}}$$($$e^m_{n\rightarrow k}(t)$, $v^m_{n\rightarrow k}$$(t))$}:
\begin{eqnarray}\nonumber
\lefteqn{\!\!\!\!\!\!\!\!\!\!\!\!\!\!\!\!\!\!\!\!\!\!\!\!\!\!\!\!\!\!\!\!\!\!\!\!\!\!\!\!\!\!\!\!\!\!\!\!\!\!\!\!\!\!\!\!\!\!\!\!\!\!\!\!\!\!
e^m_{n\rightarrow k}(t)=E[g_k|y^m_n, {\hat{\eta}}^m_{nk}, \lambda_k=1]}\\ \label{mUpSum}
\lefteqn{\!\!\!\!\!\!\!\!\!\!\!\!\!\!\!\!\!\!\!\!\!\!\!\!\!\!\!\!\!\!\!\!\!\!\!\!\!\!\!\!\!\!\!\!
=(\hat{h}^m_{n,k})^{-1}(y^m_n-e^m_{nk}(t)),}
\end{eqnarray}
\vspace{-0.6cm}
\begin{eqnarray} \nonumber
\lefteqn{\!\!\!\!\!\!\!\!\!\!\!\!\!\!\!\!\!\!\!\!\!\!\!\!\!\!\!\!\!\!\!\!\!\!\!\!\!\!\!\!\!\!\!\!\!\!\!\!\!\!\!\!\!\!\!\!\!\!\!\!\!\!\!\!\!\!
v^m_{n\rightarrow k}(t)=Var[g_k|y^m_n, {\hat{\eta}}^m_{nk}, \lambda_k=1]} \\ \label{vUpSum}
\lefteqn{\!\!\!\!\!\!\!\!\!\!\!\!\!\!\!\!\!\!\!\!\!\!\!\!\!\!\!\!\!\!\!\!\!\!\!\!\!\!\!\!\!\!\!\!
=(\hat{h}^m_{n,k})^{-2}v^m_{nk}(t),}
\end{eqnarray}
where $E[a|b]$ and $Var[a|b]$ denote the conditional expectation and variance of variable $a$ when given variable $b$ and Eq.\;(\ref{mUpSum}) and Eq.\;(\ref{vUpSum}) are derived from the fact that $\lambda_k$ and $g_k$ are independent of each other. Let initial mean~vector ${{\bm{e}}^m_n}(0)=[e^m_{1\rightarrow n}(0), ..., e^m_{K\rightarrow n}(0)]^T$ and variance~vector ${\bm{v}}^m_n(0)=[v^m_{1\rightarrow n}(0), ..., v^m_{K\rightarrow n}(0)]^T$ be $\bm{0}$ and $+\bm{\infty}$~respectively, where $\bm{0}$ and $+\bm{\infty}$ denote the vector forms of $0$ and $+ \infty$.

\emph{2) Bernoulli message update for $\lambda_k$}: 
\begin{eqnarray} \nonumber 
\lefteqn{\!\!\!\!\!\!\!\!\!\!\!\!\!\!\!\!\!\!\!\!\!\!\!\!\!\!\!\!\!\!\!\!\!\!\!\!\!\!\!\!\!\!\!\!\!\!\!\!\!\!\!\!\!\!\!\!\!\!\!\!\!\!\!\!\!\!\!\!
p^m_{n\rightarrow k}(t)=\big[{1+\frac{P(y^m_n|\lambda_k=0,{\hat{\eta}}^m_{nk})}
{P(y^m_n|\lambda_k=1,{\hat{\eta}}^m_{nk})}}\big]^{-1}} \\ \label{pUpsum}
\lefteqn{\!\!\!\!\!\!\!\!\!\!\!\!\!\!\!\!\!\!\!\!\!\!\!\!\!\!\!\!\!\!\!\!\!\!\!\!\!\!\!\!\!\!\!\!\!\!\!\!\!\!\!\!\!\!\!\!\!\!\!\!\!\!\!\!\!\!\!\!
{=\frac{1}{1+\frac{f(y^m_n,~e^m_{nk}(t),~v^m_{nk}(t))}{f(y^m_n,~\hat{h}^m_{n,k}{e}^m_{k \rightarrow n}(t)+e^m_{nk}(t),~(\hat{h}^m_{n,k})^2
{v}^m_{k \rightarrow n}(t)+v^m_{nk}(t))}}},
}
\end{eqnarray}
where $f(y, a, b)$ is the standard Gaussian probability density function (PDF) with respect to variable $y$ whose mean is $a$ and variance is $b$. 
Let initial non-zero probability vector ${{\bm{p}}^m_n}(0)=[p^m_{1\rightarrow n}(0), ..., p^m_{K\rightarrow n}(0)]^T$ be $=0.5 \times {\bm{1}}$, where $\bm{1}$ denotes the all-ones vector.
\subsection{Bernoulli-Gaussian Message Update at Variable Node}\label{BGVN}
As shown in Fig.~\ref{MessPass}(b), we present the update process for messages passing from the~$k$-th variable node to the~$n$-th sum node of~the $m$-th RRH. At first, let ${\bm{\bar{e}}}=[\bar{e}_1, ..., \bar{e}_K]^T$, ${\bm{\bar{v}}}=[\bar{v}_1, ..., \bar{v}_K]^T$ be the priori mean and variance of $\bm{g}$, and ${\bm{\bar{p}}}=[\bar{p}_1, ..., \bar{p}_K]^T$ be the priori non-zero probability of $\bm{\lambda}$. We assume that $\bar{e}_k=0$, $\bar{v}_k=\rho^{-1}$, and $\bar{p}_k=\rho$, $k \in \mathcal{K}$. Then, based on the estimated messages from all sum nodes, at the ($t+1$)-th iteration, the $k$-th variable node outputs mean $e^m_{k \rightarrow n}(t+1)$ and variance $v^m_{k \rightarrow n}(t+1)$ for $g_k$, and non-zero probability $p^m_{k \rightarrow n}(t+1)$ for $\lambda_k$, which are sent to the $n$-th sum node of the $m$-th RRH.

\emph{1) Gaussian message update for $g_k$$\sim$${\mathcal{N}}$$($$e^m_{k\rightarrow n}(t+1)$, $v^m_{k\rightarrow n}$$(t+1))$}:
According to the update rules of Gaussian message~\cite{Lei2015,Lei2016,Lei20162}, PDF of the output Gaussian message from a variable node is the normalized product of PDFs of the input Gaussian messages. Therefore, we can obtain
\begin{eqnarray} \nonumber 
\lefteqn{\!\!\!\!\!\!\!\!\!\!\!\!\!\!\!\!\!\!\!\!\!\!\!\!\!\!\!\!\!\!\!\!\!\!\!\!\!\!\!\!\!\!\!\!\!\!\!\!\!\!\!\!\!\!\!\!\!\!\!\!\!\!\!\!\!\!\!\!\!\!
v^m_{k \rightarrow n}(t+1)=Var[g_k|{\bm{V}}^k_{\setminus\{m, n\}}(t), {\bar{v}_k}]} \\ \label{VUpVa}
\lefteqn{\!\!\!\!\!\!\!\!\!\!\!\!\!\!\!\!\!\!\!\!\!\!\!\!\!\!\!\!\!\!\!\!\!\!\!\!\!\!\!\!\!\!\!\!\!\!\!\!\!\!\!\!\!\!\!\!\!\!\!\!\!\!\!\!\!\!\!\!\!\!
={\big[{\bar{v}}^{-1}_k\!\!+\!\!\!\!\sum_{\small{i\in \mathcal{M}\setminus m}}\sum_{j\in \mathcal{D}}v_{j \rightarrow k}^i(t)^{-1}\!\!+\!\!\!\sum_{d \in \mathcal{D}\setminus n}\!\!\!v_{d\rightarrow k}^m(t)^{-1}\big]^{-1},}} 
\end{eqnarray}
\vspace{-0.4cm}
\begin{eqnarray}\nonumber 
\lefteqn{\!\!\!\!\!\!\!\!\!\!\!\!\!\!\!\!\!\!\!\!\!\!\!\!\!\!\!\!\!\!\!\!\!\!\!\!\!\!\!\!\!\!\!\!\!\!\!\!\!\!\!\!\!\!\!\!\!\!\!\!\!\!\!\!\!\!\!\!\!\!
e^m_{k \rightarrow n}(t+1)=E[g_k|{\bm{V}}^k_{\setminus\{m, n\}}(t), {\bm{E}}^k_{\setminus\{m, n\}}(t), {\bar{v}_k, {\bar{e}}_k}]} \\ \label{mUpVa}
\lefteqn{\!\!\!\!\!\!\!\!\!\!\!\!\!\!\!\!\!\!\!\!\!\!\!\!\!\!\!\!\!\!\!\!\!\!\!\!\!\!\!\!\!\!\!\!\!\!\!\!\!\!\!\!\!\!\!\!\!\!\!\!\!\!\!\!\!\!\!\!\!\!
=v^m_{k \rightarrow n}(t+1)\big[\frac{\bar{e}_k}{\bar{v}_k}\!+\!\!\!\!\!\sum_{i \in \mathcal{M}\setminus m}\sum_{j\in \mathcal{D}}\frac{e_{j \rightarrow k}^i(t)}{v_{j \rightarrow k}^i(t)}\!+\!\!\!\!\sum_{d \in \mathcal{D}\setminus n}\!\!\frac{e_{d \rightarrow k}^m(t)}{v_{d\rightarrow k}^m(t)}\big],}
\end{eqnarray}
where ${\bm{V}}^k(t)$$=$$[v_{j \rightarrow k}^i(t)]_{MN \times 1}$ and ${\bm{E}}^k(t)=[e_{j \rightarrow k}^i(t)]_{MN \times 1}$ denote the mean and variance vectors associated with ${g}_k$ from all sum nodes, $i \in \mathcal{M}$, $j \in \mathcal{D}$, $\mathcal{M}$$=$$\{$$1,$ $...,$ $M\}$,
$\mathcal{D}=\{1, ..., N\}$, and $\setminus\{m, n\}$ denotes that $j \neq n$ if and only if $i=m$.

\emph{2) Bernoulli message update for $\lambda_k$}:
By combining non-zero probability ${\bm{P}}^k(t)=[p_{j \rightarrow k}^i(t)]_{MN \times 1}$ associated with ${\lambda}_k$ from all sum nodes,
where $i \in \mathcal{M}$ and $j \in \mathcal{D}$, we can obtain
\begin{eqnarray} \nonumber 
\lefteqn{\!\!\!\!\!\!\!\!\!\!\!\!\!\!\!\!\!\!\!\!\!\!\!\!\!\!\!\!\!\!\!\!\!\!\!\!\!\!\!\!\!\!\!\!\!\!\!\!\!\!\!\!\!\!\!\!\!\!\!\!\!\!\!\!\!\!\!\!\!\!
p^m_{k \rightarrow n}(t+1)=\big[{1+\frac{P(\lambda_k=0|{\bm{P}}_{\setminus\{m, n\}}(t),{\bar{p}}_k)}{P(\lambda_k=1|{\bm{P}}_{\setminus\{m, n\}}(t),{\bar{p}}_k)}}\big]^{-1}} \\ \label{PUpVa}
\lefteqn{\!\!\!\!\!\!\!\!\!\!\!\!\!\!\!\!\!\!\!\!\!\!\!\!\!\!\!\!\!\!\!\!\!\!\!\!\!\!\!\!\!\!\!\!\!\!\!\!\!\!\!\!\!\!\!\!\!\!\!\!\!\!\!\!\!\!\!\!\!\!\!
=\!\!\frac{1}
{1\!+\!
\frac{(1-\bar{p}_k) [\prod\nolimits_{i\in \mathcal{M}\setminus m}\prod\nolimits_{j\in \mathcal{D}}(1-p^i_{j \rightarrow k}(t))]\prod\nolimits_{d\in \mathcal{D}\setminus n}(1-p^m_{d \rightarrow k}(t))}
{{\bar{p}_k [\prod\nolimits_{i\in \mathcal{M}\setminus m}\prod\nolimits_{j\in \mathcal{D}}p^i_{j \rightarrow k}(t)]\prod\nolimits_{d\in \mathcal{D}\setminus n}p^m_{d \rightarrow k}(t)}}}.}
\end{eqnarray}

Considering a large amount of probability multiplications in Eq.~(\ref{PUpVa}) easily cause the storage overflow in the simulations, we transform probability calculations into log-likelihood ratio (LLR) calculations by using function $L(p)$ $=$ ${\rm{log}}\frac{p}{1-p}=-{\rm{log}}(p^{-1}-1)$. Specifically, we denote the LLR forms of $\bar{p}_k$ 
and $p^m_{k \rightarrow n}(t)$ as $\bar{{\ell}}_k=\mathcal{L}(\bar{p}_k)$ and ${\ell}^m_{k \rightarrow n}(t)=\mathcal{L}(p^m_{k \rightarrow n}(t))$ respectively. Then, Eq.~(\ref{PUpVa}) is transformed into
\begin{eqnarray}\label{LPUpVa}
\lefteqn{\!\!\!\!\!\!\!\!\!\!\!\!\!\!\!\!\!\!\!\!\!\!\!\!\!\!\!\!\!\!\!\!\!\!\!\!\!\!\!\!\!\!\!\!\!\!\!\!\!\!\!\!\!\!\!\!\!\!\!\!\!\!\!\!\!\!\!\!\!\!
{\ell}^m_{k \rightarrow n}(t+1)=\bar{{\ell}}_k\!+\!\!\!\sum_{i\in \mathcal{M}\setminus m}\sum_{j\in \mathcal{D}}{\ell}^i_{j \rightarrow k}(t)\!+\!\!\sum_{d\in \mathcal{D}\setminus n}\!\!{\ell}^m_{d \rightarrow k}(t),}
\end{eqnarray}
where initialization of ${{\bm{L}}^m_n}(0)$\;$=$\;$\mathcal{L}({{\bm{p}}^m_n}(0))$ is $\bm{0}$. Correspondingly, input message ${p}^m_{k\rightarrow n}(t)$ of Eq.~(\ref{ETAM}) and Eq.~(\ref{ETAV}) is equal to $[{{\rm{tanh}}(\frac{{\ell}^{m}_{{k} \rightarrow {n}}(t)}{2})+1}]/{2}$.
\subsection{Decision and Output of BGMP}
The BGMP algorithm is performed iteratively between the sum nodes and variable nodes, where Eq.~(\ref{mUpSum})--Eq.~(\ref{pUpsum}) are the update functions for messages at sum nodes, and Eq.~(\ref{VUpVa})-- Eq.~(\ref{LPUpVa}) are the update functions for messages at variable nodes. The whole iterative process is stop until when the preset maximum iterative number is reached or the MSE requirement is satisfied. According to the message passing rules~\cite{LDPC,Lei2015}, the decision depends on the full messages at the Gaussian and Bernoulli nodes which combine priori messages and input messages from sum nodes together. The full messages of mean and variance of $\bm{g}$ are denoted as $\bm{\tilde{e}}$ and $\bm{\tilde{v}}$ respectively, and those of non-zero probability and the corresponding LLR of $\bm{\lambda}$ are denoted as $\bm{\tilde{{P}}}$ and $\bm{\tilde{{\ell}}}$ respectively. The $k$-th entries
of $\bm{\tilde{e}}$, $\bm{\tilde{v}}$, $\bm{\tilde{{P}}}$, and $\bm{\tilde{{\ell}}}$, $k\in \mathcal{K}$, are
\begin{eqnarray} \label{tlv}
\lefteqn{\!\!\!\!\!\!\!\!\!\!\!\!\!\!\!\!\!\!\!\!\!\!\!\!\!\!\!\!\!\!\!\!\!\!\!\!\!\!\!\!\!\!\!\!\!\!\!\!\!\!\!\!\!\!\!\!\!\!\!\!\!\!\!\!\!\!\!\!\!\!
\tilde{v}_k=[\bar{v}_k^{-1}+\sum_{i\in \mathcal{M}}\sum_{j\in \mathcal{D}} v^i_{j\rightarrow k}(t)^{-1}]^{-1},} \\ \label{tlm}
\lefteqn{\!\!\!\!\!\!\!\!\!\!\!\!\!\!\!\!\!\!\!\!\!\!\!\!\!\!\!\!\!\!\!\!\!\!\!\!\!\!\!\!\!\!\!\!\!\!\!\!\!\!\!\!\!\!\!\!\!\!\!\!\!\!\!\!\!\!\!\!\!\!
\tilde{e}_k=\tilde{v}_k[{\bar{e}_k}{\bar{v}_k}^{-1}+\sum_{i\in \mathcal{M}}\sum_{j\in \mathcal{D}}{e_{j \rightarrow k}^i(t)}{v_{j \rightarrow k}^i(t)}^{-1}],}\\ \label{tlL}
\lefteqn{\!\!\!\!\!\!\!\!\!\!\!\!\!\!\!\!\!\!\!\!\!\!\!\!\!\!\!\!\!\!\!\!\!\!\!\!\!\!\!\!\!\!\!\!\!\!\!\!\!\!\!\!\!\!\!\!\!\!\!\!\!\!\!\!\!\!\!\!\!\!
\tilde{\ell}_k=\bar{\ell}_k+\sum_{i\in \mathcal{M}}\sum_{j\in \mathcal{D}}{\ell}^i_{j \rightarrow k}(t),
}\\
\label{tlp}
\lefteqn{\!\!\!\!\!\!\!\!\!\!\!\!\!\!\!\!\!\!\!\!\!\!\!\!\!\!\!\!\!\!\!\!\!\!\!\!\!\!\!\!\!\!\!\!\!\!\!\!\!\!\!\!\!\!\!\!\!\!\!\!\!\!\!\!\!\!\!\!\!\!
\tilde{p}_k=[{{\rm{tanh}}(\frac{\tilde{\ell}_k}{2})+1}]/{2},
}
\end{eqnarray}
Based on Eq.~(\ref{tlv})--Eq.~(\ref{tlp}), the $k$-th entry of final estimation $\bm{\tilde{\lambda}}$ of $\bm{\lambda}$ is
\begin{equation} \nonumber
\tilde{\lambda}_k = \left\{ {\begin{array}{*{20}c}
{1,\quad {\rm{when}}\;\tilde{\ell}_k > 0}, \\
{0,\quad {\rm{when}}\;\tilde{\ell}_k \leq 0},
\end{array}} \right.
\end{equation}
and the final estimation $\bm{\tilde{x}}$ of $\bm{x}$ is $\bm{\tilde{x}}= \bm{\tilde{\lambda}} \circ \bm{\tilde{{P}}}\circ \bm{\tilde{e}}$.
\subsection{Complete BGMP Algorithm}
Now we present the complete process of BGMP algorithm. Assume $i\in\mathcal{M}$, $j\in\mathcal{D}$, and $k\in\mathcal{K}$.
Let $\bm{\hat{H}}$\;$=$\;$[\bm{\hat{H}}^1,$ $...,$ $\bm{\hat{H}}^M]^T$\;$=$\;$[\hat{h}_{j,k}^i]_{MN\times K}$ be the whole matrix. We define $\mathcal{J}(i)$ as the set of neighbors of the $i$-th node, which denotes that there is a edge connecting the $i$-th node
and any $d$-th node, $d$\;$\in$\;$\mathcal{J}(i)$, i.e., $\hat{h}_{j, d}^i$\;$\neq$\;$0$. Moreover, let $\bm{y}$ $=$ $[y_{j}^i]_{MN\times \rm{1}}$,
$\bm{E^{\eta}}(t)$ $=$ $[e^m_{nk}(t)]_{MN\times K}$, $\bm{V^{\eta}}(t)$ $=$ $[v^m_{nk}(t)]_{MN\times K}$,~$\bm{\sigma^{\eta}}$$=$$[\sigma_{mn}^2]_{MN \times \rm{1}}$,
$\bm{E}^s(t)=[e_{j \rightarrow k}^i(t)]_{MN \times K}$, $\bm{V}^s(t)$$=$$[v_{j \rightarrow k}^i(t)]_{MN \times K}$,~$\bm{P}^s(t)$$=$$[p_{j \rightarrow k}^i(t)]_{MN \times K}$,
$\bm{L}^s(t)$ $=$ $[\ell_{j \rightarrow k}^i(t)]_{MN \times K}$, $\bm{{E}}^v(t)$\;$=$\;$[e_{k \rightarrow j}^i(t)]_{K\times MN}$, $\bm{V}^v(t)$ $=$ $[v_{k \rightarrow j}^i(t)]_{K\times MN}$,
$\bm{P}^v(t)$ $=$ $[p_{k \rightarrow j}^i(t)]_{K\times MN}$, $\bm{L}^v(t)$ $=$ $[\ell_{k \rightarrow j}^i(t)]_{K\times MN}$.
The output of function ${\text{sign}}(a)$ is equal to $1$ when $a>0$ and $0$ when $a\leq 0$. The complete BGMP algorithm is given in Algorithm~1.
\begin{algorithm}
\caption{Bernoulli-Gaussian Message Passing (BGMP)}
\begin{algorithmic}[1]
\State {\small{\textbf{Input:} {$\bm{y}$, {$\bm{\hat{H}}$, $\bm{\sigma^{\eta}}$, ${\bm{\bar{e}}}$, ${\bm{\bar{v}}}$, ${\bm{\bar{p}}}$, and $\rho \!\in\! (0,1)$}}.
\State \textbf{Initialization:} $t=0$, $\mathbf{E}^v(\!0\!)=\mathbf{0}$, $\mathbf{V}^v(\!0\!)\!=\!+\boldsymbol{\infty}$, and $\bm{L}^v(\!0\!)\!=\!\bm{\!0}$.
\State \textbf{Repeat:}  set $t \Leftarrow t+1$,
\State {\textbf{for} $i=1,... MN$, $k \in \mathcal{J}(i)$, \textbf{do}\\
       $\bm{P}^v_{k,i}(t)=[{\rm{tanh}}(\bm{L}^v_{k,i}(t)/2)+{1}]/2$,\\
        \textbf{end}
\State \textbf{for} $i=1,... MN$, $k \in \mathcal{J}(i)$, \textbf{do}
\State  Define $U_i=\sum_{k}U_{i,k}=\sum_{k}\bm{\hat{H}}_{i,k}\bm{P}^{v}_{k,i}(t)\bm{E}^v_{k,i}(t)$, $W_i=$
$\sum_{k}W_{i,k}$$=$$\sum_{k}(\bm{\hat{H}}_{i,k})^2\bm{P}^{v}_{k, i}(t)$$\big[(\bm{V}^v_{k, i}(t)$$+$$(1$$-\bm{P}^{v}_{k,i}(t))\bm{E}^v_{k, i}(t)^2\big]$,
\State \[ \!\!\!\!\!\!\!\!\begin{array}{c}
\left[\!\!\!\! \begin{array}{c}
\bm{E^{\eta}}_{i,k}(t)\\
\bm{V^{\eta}}_{i,k}(t)
\end{array} \!\!\!\!\right]\!\! \!=\!\!\! \left[\! \!\!\!\begin{array}{c}
U_i-U_{i,k},\\
W_i-W_{i,k} +\bm{\sigma^{\eta}}_{i, k}
\end{array} \!\!\!\!\!\right]\!,
\end{array}\]
\State \vspace{-0.4cm}\[ \!\!\!\!\!\!\!\!\begin{array}{c}
\left[\!\!\!\! \begin{array}{c}
\bm{E^{s}}_{i,k}(t)\\
\bm{V^{s}}_{i,k}(t)\\
\bm{L^{s}}_{i,k}(t)\\
\end{array} \!\!\!\!\right]\!\! \!=\!\!\! \left[\! \!\!\!\begin{array}{c}
(\bm{\hat{H}}_{i,k})^{-1}(\bm{Y}_i-\bm{E^{\eta}}_{i,k}(t))\\
(\bm{\hat{H}}_{i,k})^{-2}\bm{V^{\eta}}_{i,k}(t)\\
{{\!-\!\frac{1}{2}}}\!\log{\!\big[\!{1\!\!+\!\! \tfrac{\bm{\hat{H}}_{\!i,k\!}^2\bm{V}^v_{\!k,i\!}(\!t\!)} {\bm{V^{\eta}}_{\!i,k\!}(\!t\!)}\!\big]} \!\!+\!\!\tfrac{\bm{\hat{H}}_{\!i,k\!}^2\bm{V}^v_{\!k,i\!}(\!t\!)({\bm{y}}_{\!i\!}-
\bm{E^{\eta}}_{i,k}(t))^2}
{2{\bm{V^{\eta}}_{\!i,k\!}(\!t\!)}[{\bm{V^{\eta}}_{\!i,k\!}(\!t\!)}+
\bm{\hat{H}}_{\!i,k\!}^2\bm{V}^v_{\!k,i\!}(\!t\!)]}+\;\;} \\
{\frac{\bm{\hat{H}}_{\!i,k\!}\bm{E}^v_{\!k,i\!}(\!t\!)\bm{V^{\eta}}_{\!i,k\!}(\!t\!)\big[
2({\bm{y}}_{\!i\!}-\bm{E^{\eta}}_{\!i,k\!}(\!t\!))-\bm{\hat{H}}_{\!i,k\!}\bm{E}^v_{\!k,i\!}(\!t\!)
\big]} {2{\bm{V^{\eta}}_{\!i,k\!}(\!t\!)}[{\bm{V^{\eta}}_{\!i,k\!}(\!t\!)}+
\bm{\hat{H}}_{\!i,k\!}^2\bm{V}^v_{\!k,i\!}(\!t\!)]}}
\end{array} \!\!\!\!\!\right]\!,
\end{array}\]\\
        \vspace{-0.1cm}\textbf{end}
\State \textbf{for} $k=1, ..., K$, $i \in \mathcal{J}(k)$, \textbf{do}
\State Define $V_k=\sum_{i}V_{k,i}=\sum_{i}\bm{V^{s}}_{i,k}(t)^{-1}$, $E_k=\sum_{i}E_{k,i}=\sum_{i}{\bm{E^{s}}_{i,k}(t)}{\bm{V^{s}}_{i,k}(t)}^{-1}$, $L_k=\sum_{i}L_{k,i}=\sum_{i}\bm{L}^s_{i,k}(t)$,
\State \vspace{-0.1cm}\[ \!\!\!\!\!\!\!\!\begin{array}{c}
\left[\!\!\!\! \begin{array}{c}
\bm{V^{v}}_{k,i}(t+1)\\
\bm{E^{v}}_{k,i}(t+1)\\
\bm{L^{v}}_{k,i}(t+1)\\
\end{array} \!\!\!\!\right]\!\! \!=\!\!\! \left[\! \!\!\!\begin{array}{c}
\big[(\bm{\bar{v}}_{k,i})^{-1}+ V_k-V_{k,i}\big]^{-1}\\
\bm{V^{v}}_{k,i}(t+1)\big[{\bm{\bar{e}}_k}{\bm{\bar{v}}_k}^{-1}+E_k-E_{k,i}
\big]\\
\bm{\bar{\ell}}_{k,i}+L_k-L_{k,i}
\end{array} \!\!\!\!\!\right]\!,
\end{array}\]\\
        \vspace{-0.1cm}\textbf{end}
\State \textbf{Until:} stopping criteria are satisfied
\State \textbf{for} $k=1, ..., K$, \textbf{do}
\State \vspace{-0.1cm}\[ \!\!\!\!\!\!\!\!\begin{array}{c}
\left[\!\!\!\! \begin{array}{c}
\bm{\tilde{v}}_k\\
\bm{\tilde{e}}_k\\
\bm{\tilde{\ell}}_k\\
\bm{\tilde{p}}_k\\
\end{array} \!\!\!\!\right]\!\! \!=\!\!\! \left[\! \!\!\!\begin{array}{c}
\big[(\bm{\bar{v}}_k)^{-1}+ \sum_{i \in \mathcal{J}(k)} \bm{V^{s}}_{i,k}(t)^{-1}\big]^{-1}\\
\bm{\tilde{v}}_k \big[{\bm{\bar{e}}_k}/{\bm{\bar{v}}_k}^{-1}+\sum_{j \in \mathcal{J}(k)}{\bm{E^{s}}_{j,k}(t)}{\bm{V^{s}}_{j,k}(t)}^{-1}
\big]\\
\bm{\bar{\ell}}_k+\sum_{j \in \mathcal{J}(k)}\bm{L}^s_{j,k}(t)\\
{[\rm{tanh}({\bm{\tilde{\ell}}_k}/{2})+1]}/{2},
\end{array} \!\!\!\!\!\right]\!,
\end{array}\]\\
        \vspace{-0.1cm}\textbf{end}
\State \textbf{Output:} \;\;$\bm{\tilde{\lambda}}=\rm{sign}(\bm{\tilde{\ell}})$, \rm{and} $\bm{\tilde{x}}= \bm{\tilde{\lambda}} \circ \bm{\tilde{{P}}}\circ \bm{\tilde{e}}$
}}
}
\end{algorithmic}
\end{algorithm}
\section{Numerical Results}
In this section, we investigate the performance of the proposed BGMP algorithm over the grant-free C-RAN system. We assume that the RRHs and users are uniformly located over a square network whose side length is $5~\text{km}$. The path loss exponent $\alpha$ is $2.25$. The number of RRHs and users is $M=120$ and $K=200$, where each RRH has $N=10$ antennas and the probability of user activity $\rho=0.3$. The maximum iteration number of the BGMP
algorithm is $\tau_{\text{max}}=50$. The average receive signal-to-noise ratio (RSNR) is $\frac{PE[\sum_{m \in \mathcal{M}} ||\bm{H}^m||^2_2]}{MN\sigma_n^2}$. We measure the performances in terms of the average MSE and user state error (USE), i.e., $\text{MSE}=\frac{1}{K}E[||\bm{x}-\bm{\tilde{x}}||^2_2]$ and $\text{USE}=\frac{1}{K}E[||\bm{\lambda}-\bm{\tilde{\lambda}}||_1]$.
\subsection{Benchmark Detections}
To evaluate recovery accuracy of the BGMP algorithm, we present three benchmark detections: genie-aided minimum mean-square error (GA-MMSE), genie-aided sparse MMSE (GA-SMMSE), and general SMMSE, where the genie aid denotes that the detector knows the non-zero locations of $\bm{x}$ in advance and the sparseness denotes that the detector exploits sparsified matrix $\bm{\hat{H}}$ instead of original matrix $\bm{H}$. The estimations of these detectors are
\[\!\!\!\!\!\!\!\!\!\!\!\!\!\!\!\!\!\!\!\!\!\!\!\!\!\!
\bm{x}^{\rm{GA-MMSE}}_{\setminus\{0\}}=\big(\bm{H}_{\setminus \{0\}}^T\bm{H}_{\setminus \{0\}}+\sigma_n^{2}\rho \bm{I}\big)^{-1}\bm{H}_{\setminus \{0\}}^T\bm{Y},\]
\[\!\!\!\!\!\!\!\!\!\!\!\!\!\!\!\!\!
\bm{x}^{\rm{GA-SMMSE}}_{\setminus\{0\}}=\big(\bm{\hat{H}}_{\setminus \{0\}}^T\bm{\hat{H}}_{\setminus \{0\}}+\rho \bm{\sigma^{\eta}}_{\setminus \{0\}}\bm{I}\big)^{-1}\bm{\hat{H}}_{\setminus \{0\}}^T\bm{Y},\]
\[\!\!\!\!\!\!\!\!\!\!\!\!\!\!\!\!\!\!\!\!\!\!\!\!\!\!\!\!\!\!\!\!\!\!\!\!\!\!\!\!\!\!\!\!\!\!\!\!\!\!\!\!\!\!\!\!
\bm{x}^{\rm{SMMSE}}=\big(\bm{\hat{H}}^T\bm{\hat{H}}+\rho\bm{\sigma^{\eta}}\bm{I}\big)^{-1}\bm{\hat{H}}^T\bm{Y}, \]
where\;$\setminus \{0\}$\;denotes that the entries with respect to the zeros locations of $\bm{x}$ are excluded. Since the GA-MMSE and GA-SMMSE exactly know the non-zeros locations of $\bm{x}$, they can provide the ideal limit and lower-bound performances respectively. In contrast, the SMMSE only makes use of $\bm{\hat{H}}$ so that it provides the upper-bound performance.
\subsection{Complexity Comparison}
Note that the realizable SMMSE detector has a computational complexity of $\mathcal{O}(K^3+K^2MN+KMN)$. In contrast, we discuss the computational complexity of the BGMP.  By defining channel sparsity $\it{\gamma}$ of $\bm{\hat{H}}$ as the ratio of the number of non-zero entries to that of all entries,
the number of edges in the factor graph is ${\it{\gamma}}MNK$. In each iteration,  the message update along one edge requires 
$20$ multiplication/division, and $2$ exponent/logarithm operations. Therefore, the computational complexity of the proposed BGMP is
$\mathcal{O}({\it{\gamma}}MNK\tau_{\text{max}})$. As $\it{\gamma}$ decreases, the BGMP can achieve very low complexity.
\subsection{MSE Performance Comparison}\label{MSEComp}
\begin{figure}[t!]\vspace{-0.2cm}
\centering
\includegraphics[width=0.9\columnwidth]{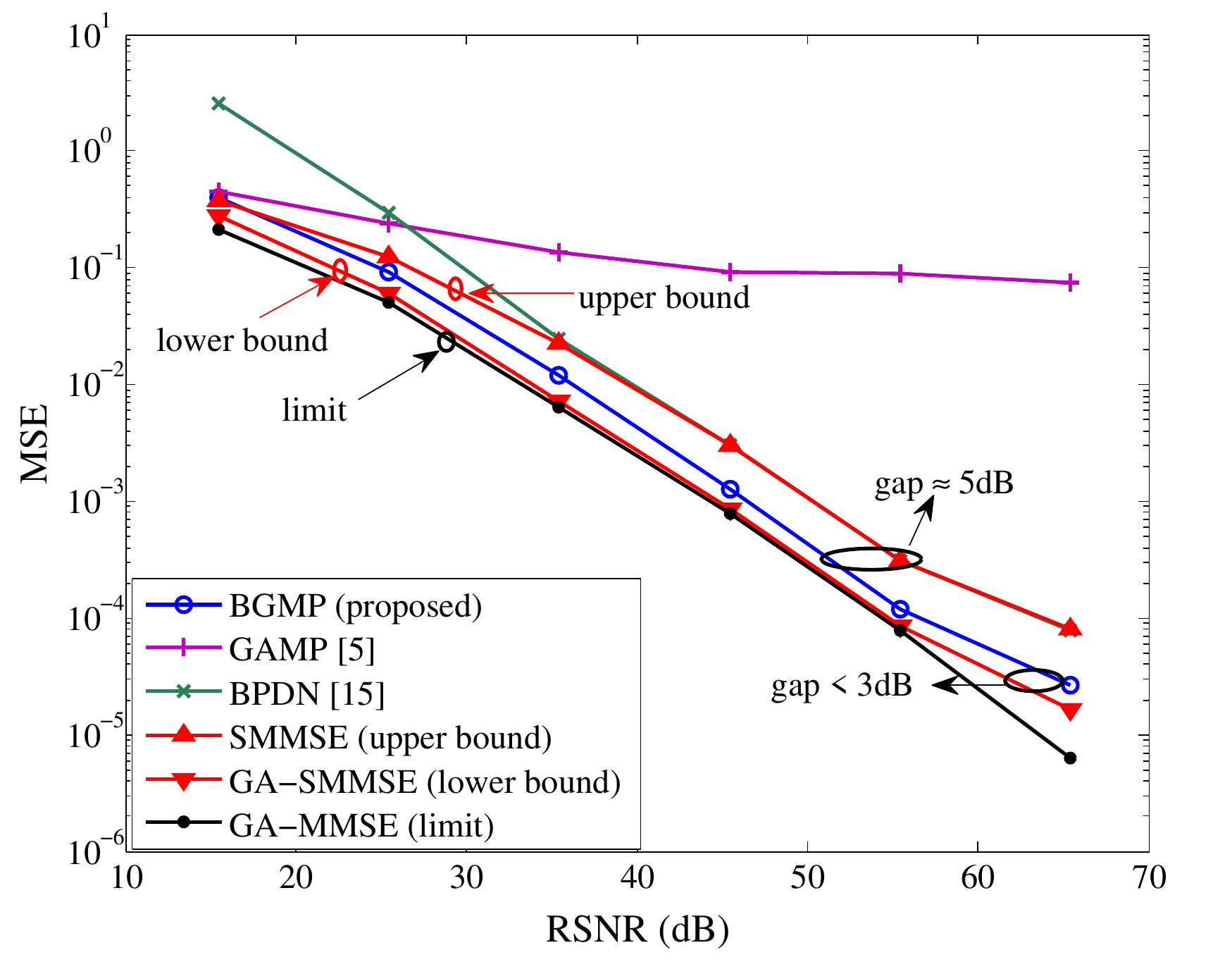}\\ \vspace{-0.3cm}
\caption{MSE Curves of the proposed BGMP, GA-MMSE, GA-SMMSE, SMMSE, GAMP~\cite{GAMP} and BPDN~\cite{BPDN}. GA-MMSE, GA-SMMSE, and SMMSE provide the limit, lower-bound, and upper-bound performances respectively. The MSE of the proposed BGMP approaches that of the GA-SMMSE at available RSNRs and has performance loss just within $3$ dB at high RSNRs.}\label{MSE} 
\end{figure}

In Fig.~\ref{MSE}, we give the MSEs of the proposed BGMP, GA-MMSE, GA-SMMSE, and SMMSE over the C-RAN system, where distance threshold $d_0=3.5~\text{km}$ and channel sparsity  ${\it{\gamma}}=0.7$. Note that the MSE curve of the BGMP is very close to that of GA-MMSE and GA-SMMSE at the entire range of RSNRs and the gap between the MSE curves of the BGMP and GA-SMMSE is  less than $3$~dB at the high RSNRs. Compared with the SMMSE, the BGMP has about $5$dB performance gain. Moreover, we also present the MSEs of GAMP~\cite{GAMP} and basis pursuit de-noising (BPDN)~\cite{BPDN}. It is noticed that the GAMP still achieves high MSE even when the RSNR is high and MSE of BPDN just converges that of SMMSE, where $\tau_{\text{max}}$ for GAMP and BPDN is also $50$. 
\subsection{BGMP Convergence and USE Performance}
\begin{figure}[h!]\vspace{-0.26cm}
\centering
\includegraphics[width=0.9\columnwidth]{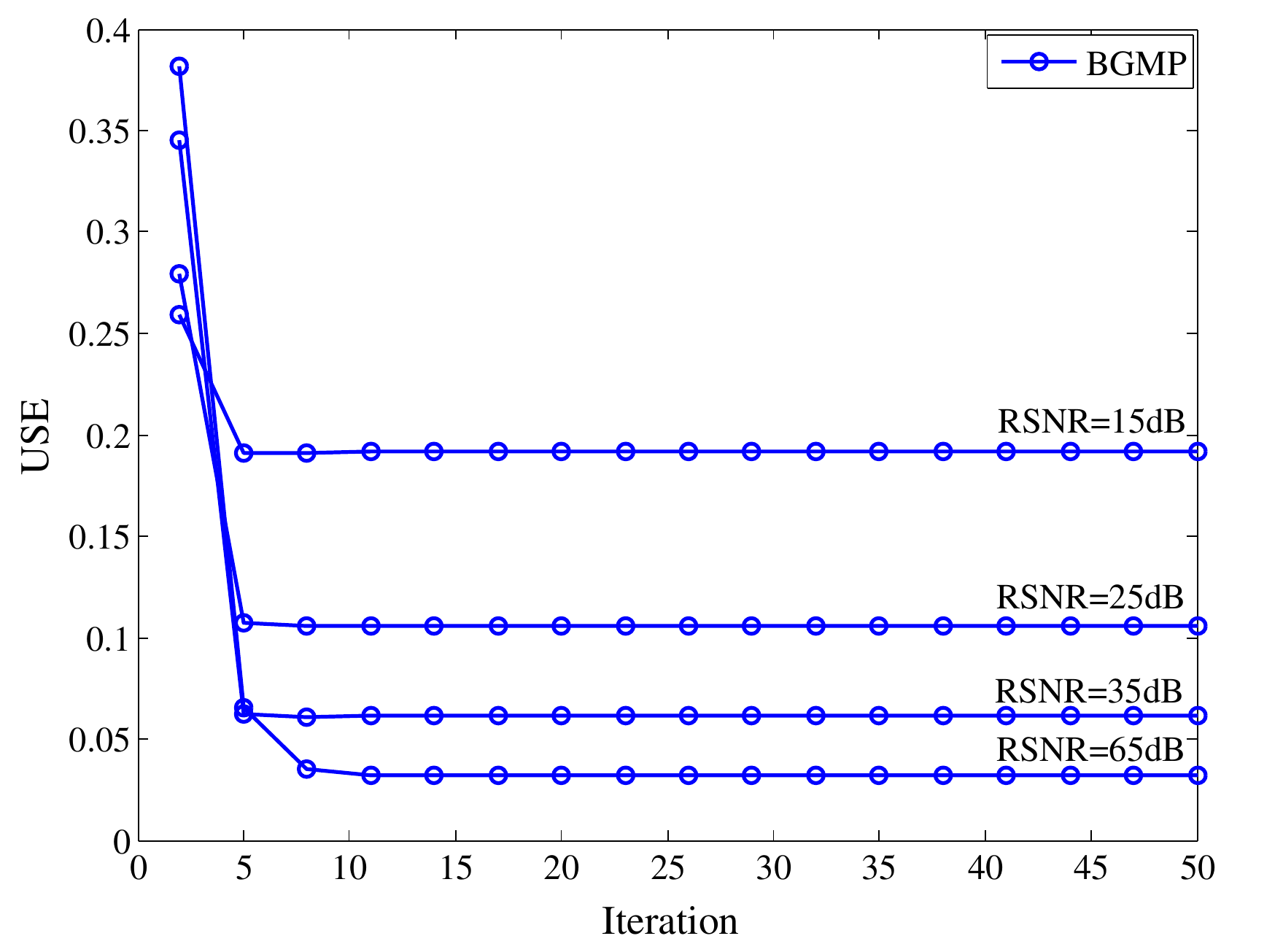}\\ \vspace{-0.3cm}
\caption{USE curves of the proposed BGMP with different RSNRs and iterations. For each RSNR, the BGMP only takes less iterations to converge.}\label{BER} 
\end{figure}

Fig.~\ref{BER} shows that USEs of the BGMP with different RSNRs and iterations, where the simulation conditions are the same as in section \ref{MSEComp}. Note that for each RSNR, the BGMP only takes less than $12$ iterations to converge, which illustrates the fast convergence of the BGMP. Additionally, as the RSNR increases, the USE of the BGMP is as low as $3\times 10^{-2}$.
\subsection{Effect of Channel Sparsity}
\begin{figure}[t!]\vspace{-0.2cm}
\centering
\includegraphics[width=0.9\columnwidth]{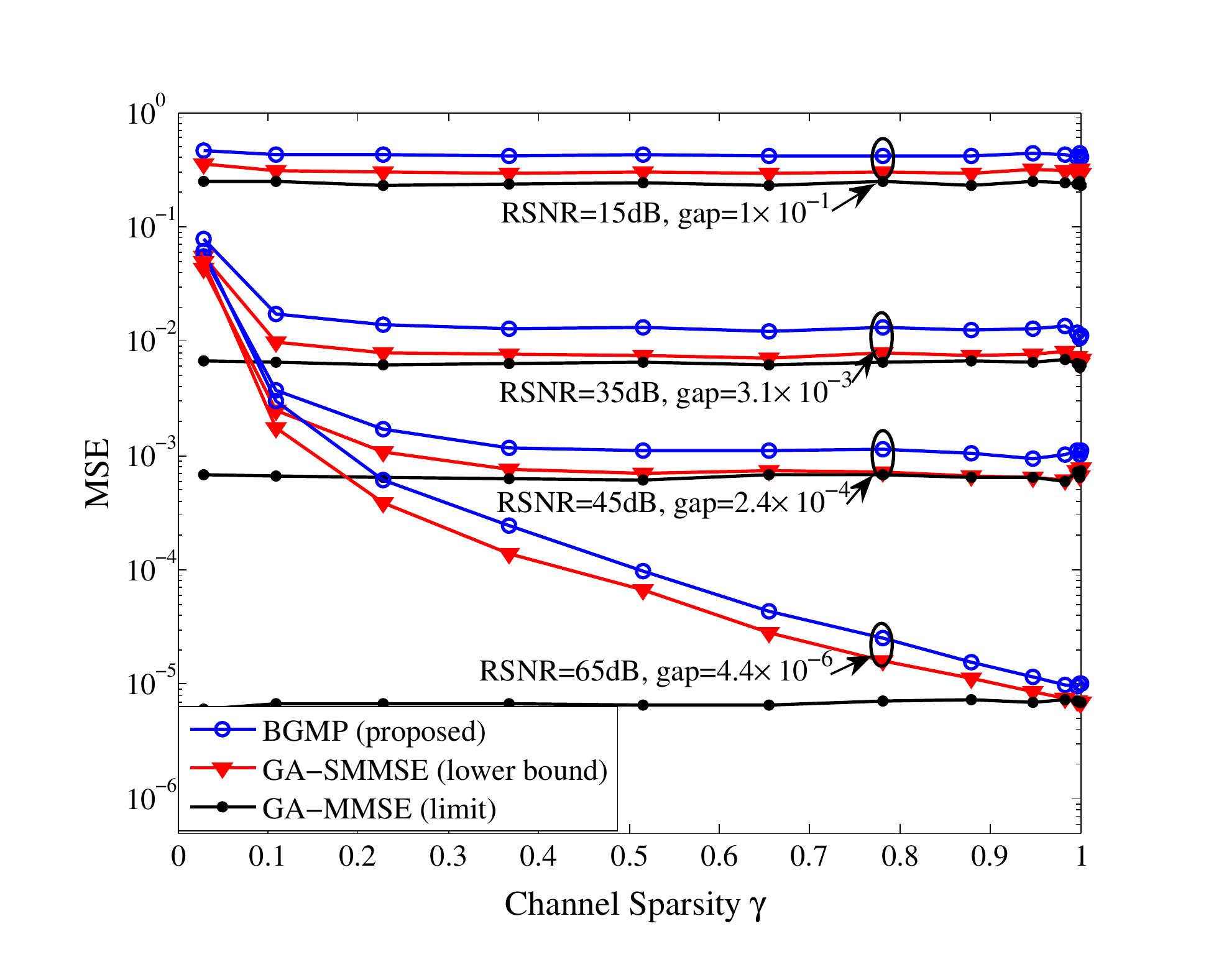}\\ \vspace{-0.3cm}
\caption{MSE Curves of the proposed BGMP, GA-MMSE (limit) and GA-SMMSE (low bound) over the sparsified C-RAN with different channel sparsity $\it{\gamma}$ and RSNRs. For each RNSR, MSE curves of the proposed BGMP approach that of GA-SMMSE at the entire range of $\it{\gamma}$.}\label{SMSE} 
\end{figure}

To investigate the robustness of the BGMP, in Fig.~\ref{SMSE} we provide the MSEs of the BGMP, GA-MMSE, and GA-SMMSE over the C-RAN with different channel sparsity and RSNRs. By directly changing the values of $d_0$, $\it{\gamma}$ changes from a small value near $0$ to $1$ accordingly. Fig.~\ref{SMSE} shows that for each RSNR, the MSE of the BGMP is close to that of GA-SMMSE at the entire range of $\it{\gamma}$. In addition, as the RSNR increases, the gaps between the MSE curves of GA-MMSE and GA-SMMSE become large. The reason is that eligible $d_0$ increases with the RNSR~\cite{Fan_Dynamic2016}, which results in
the increases of eligible $\it{\gamma}$.

Moreover, we present the USEs of the BGMP with different $\it{\gamma}$ and RSNRs. Fig.~\ref{SMSE} illustrates that for a given RSNR, $\it{\gamma}$ almost does not affect the USE performance for the BGMP.

\begin{figure}[h!]\vspace{-0.1cm}
\centering
\includegraphics[width=0.85\columnwidth]{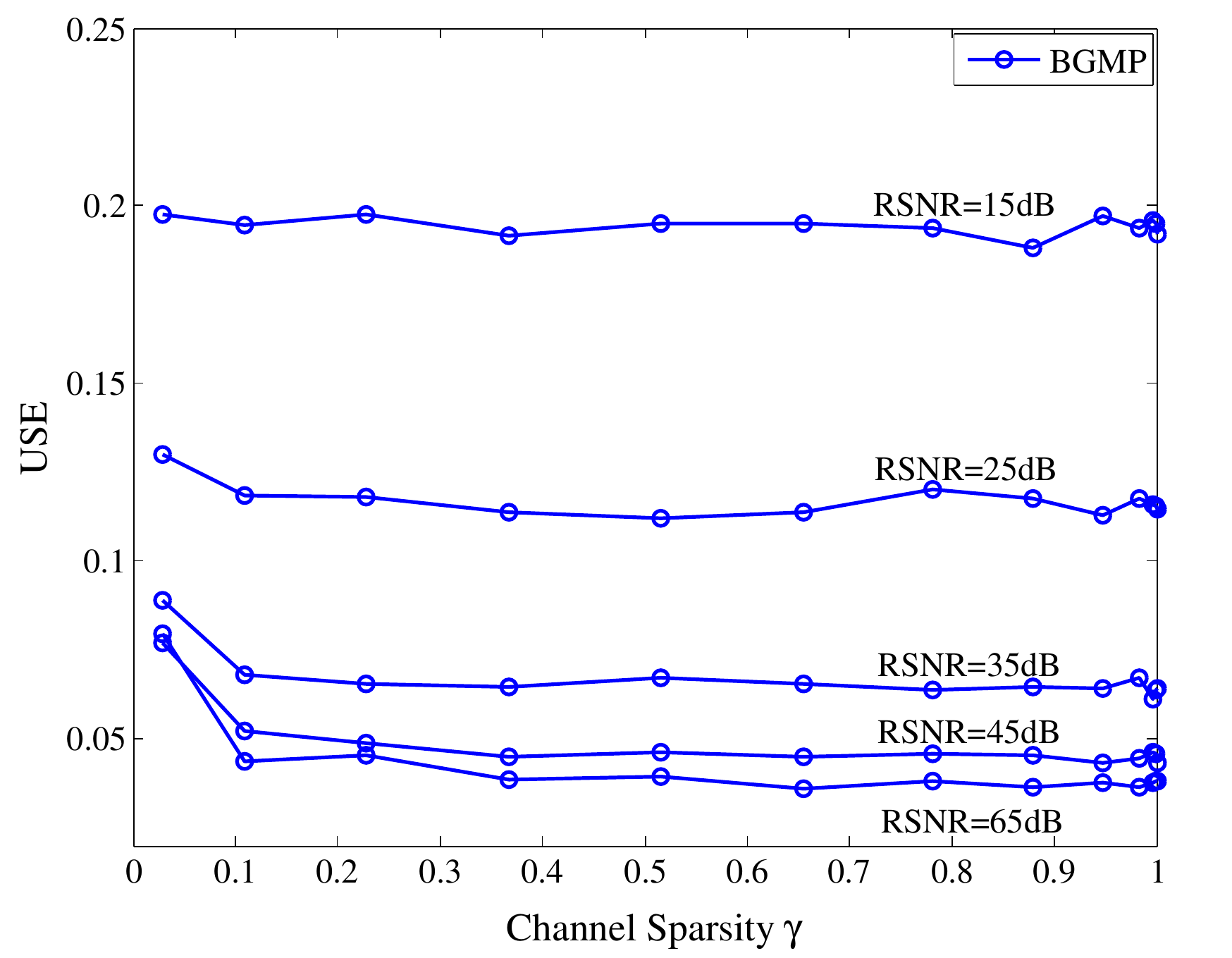}\\ \vspace{-0.25cm}
\caption{USE curves of the proposed BGMP with different channel sparsity $\it{\gamma}$ and RSNRs. For each RSNR, the USE performances of the BGMP are robust to $\it{\gamma}$ which changes from a value near $0$ to $1$.}\label{SBER} \vspace{-0.45cm}
\end{figure}

Furthermore, in  the random network, the distance between each user and each RRH is different so that the variances of entries of $\bm{\hat{H}}$
are different. As a result, the entries of $\bm{\hat{H}}$ are independent but differently distributed. Fig.~\ref{SMSE} and Fig.~\ref{SBER} further
verify that the BGMP is robust to the statistical distribution of channel.
\section{Conclusion}
In this paper, we proposed a low-complexity Bernoulli-Gaussian message passing (BGMP) algorithm for the grant-free C-RAN system. Based on the sparsified channel, the BGMP could jointly detect user activity and signal with low complexity. Numerical results showed that for different sparsified channels, the BGMP took less iterations to approach the MSE of the GA-SMMSE and low USEs. In the future work, we will provide the convergence analysis for the BGMP.


\begin{thebibliography}{h}
\bibitem{C-ran}
``C-RAN: The road towards green RAN,'' China Mobile Res. Inst., Beijing, China, Oct. 2011, White Paper, ver. 2.5.
\bibitem{FanMaga}
C.~Fan,~Y.~Zhang,~and~X.~Yuan, ``Advances and challenges toward a scalable cloud radio access network'', \emph{IEEE Commun. Magazine}, vol.~54, no. 6, pp.~29-35, Jun.~2016.
\bibitem{Zuo}
J. Zuo, J. Zhang, C. Yuen, W. Jiang, and W. Luo, ``Energy efficient user association for cloud radio access networks'', \emph{IEEE Access}, vol.~4, pp. 2429-2438, May 2016.
\bibitem{Loeliger2006}
H.~A.~Loeliger,~J.~Dauwels,~J.~Hu,~S.~Korl,~L.~Ping,~and~F.~R.~Kschischang, ``The factor graph approach to model-based signal processing,'' \emph{Proc. IEEE}, vol.~95, no.~6, pp.~1295-1322,~Jun.~2007.
\bibitem{Donoho2009}
D. L. Donoho, A. Maleki, and A. Montanari, ``Message passing algorithms for compressed sensing," \emph{Proceedings of the National Academy of Sciences}, 2009.
\bibitem{GAMP}
S.~Rangan, ``Generalized approximate message passing for estimation with random linear mixing," in \emph{Proc. IEEE ISIT}, Aug. 2011.
\bibitem{Chongwen}
C. Huang, L. Liu, C. Yuen, and S. Sun, ``A LSE and sparse message passing-based channel estimation for mmWave MIMO systems,'' in \emph{Proc. IEEE GLOBECOM Workshops}, Dec.~2016.
\bibitem{Fan_Dynamic2016}
C.~Fan,~Y.~Zhang,~and~X.~Yuan, ``Dynamic nested clustering for parallel PHY-layer processing in Cloud-RANs'', \emph{IEEE Tran. Wireless Commun.}, vol.~15, no.~3, pp.~1881-1894, Mar.~2016.
\bibitem{Lei2015}
L. Liu, C. Yuen, Y. L. Guan, Y. Li, and Y. Su, ``A low-complexity Gaussian message passing iterative detection for massive MU-MIMO systems," in \emph{Proc. IEEE ICICS}, Dec. 2015.
\bibitem{Lei2016}
L. Liu, C. Yuen, Y. L. Guan, Y. Li, and Y. Su, ``Convergence analysis and assurance Gaussian message passing iterative detection for massive MU-MIMO systems," \emph{IEEE Trans. Wireless Commun.}, vol. 15, no. 9, pp. 6487-6501, Sept. 2016.
\bibitem{Lei20162}
L. Liu, C. Yuen, Y. L. Guan, Y.~Li and C.~Huang, ``Gaussian message passing iterative detection for MIMO-NOMA systems with massive users," in \emph{Proc. IEEE GLOBECOM}, Dec. 2016.
\bibitem{Fan2015}
C.~Fan,~Y.~Zhang,~and~X.~Yuan,``Scalable uplink processing via sparse message passing in C-RAN'', in \emph{Proc. IEEE GLOBECOM}, Dec.~2015
\bibitem{Fan2016}
C.~Fan,~X.~Yuan, and~Y.~Zhang,``Randomized Gaussian message passing for scalable uplinke signal processing in C-RANs'', in \emph{Proc. IEEE ICC}, May 2016.
\bibitem{XXu2015}
X.~Xu,~X.~Rao,~and~V.~K.~N.~Lau, ``Active user detection and channel estimation in uplink CRAN systems," in \emph{Proc. IEEE ICC,} Jun. 2015.
\bibitem{Utkovski2017}
Z.~Utkovski,~O.~Simeone,~T.~Dimitrova,~and~P.~Popovski, ``Random access in C-RAN for user activity detection with limited-capactiy fronthaul'', \emph{IEEE Signal Process. Lett.}, vol. 24, no.1, pp. 17-21, Jan. 2017.
\bibitem{LDPC}
T.~J.~Richardson,~and~R.~L.~Urbanke, ``The capacity of low-density parity-check codes under message-passing decoding'', \emph{IEEE Tran. Inf. Theroy}, vol.~47, no.~2, pp. 599-618, Feb.~2001.
\bibitem{BPDN}
S.~S.~Chen,~D.~L.~Donoho,~and~M.~A.~Saunders,``Atomic decomposition by basis pursuit," \emph{SIAM J. Scientif. Comput.}, vol.~20, no.~1, pp.~33-61,~1998.
\bibitem{RandomNet}
D.~Moltchanov, ``Distance distributions in random networks," \emph{Ad Hoc Netw.}, vol.~10, no.~6, pp. 1146-1166, Mar.~2012.




\end{thebibliography}
\end{document}